\documentclass[fleqn,usenatbib]{mnras}

\usepackage{graphicx}	
\usepackage{amsmath}	
\usepackage{natbib} 
\usepackage{upgreek}
\usepackage{xcolor}
\usepackage{gensymb}
\newcommand{\cgs}{ erg\,cm$^{-2}$\,s$^{-1}$ }

\newcommand{\integral}{{\it INTEGRAL}}

\newcommand{\maxi}{{\it MAXI/GSC}}

\newcommand{\source}{MAXI J1631$-$479}

\newcommand{\nus}{{\em NuSTAR}}

\title[MAXI J1631$-$479 during the 2019 outburst]{Evolution of MAXI J1631$-$479 during the January 2019 outburst observed by \integral\//IBIS }

\author[Fiocchi et al.]{
M. Fiocchi,$^{1}$\thanks{E-mail: mariateresa.fiocchi@inaf.it }
F. Onori,$^{1}$
A. Bazzano,$^{1}$
A.J. Bird,$^{2}$
A. Bodaghee,$^{3}$
P. A. Charles,$^{2}$
\newauthor
V. A. Lepingwell,$^{2}$
A. Malizia,$^{4}$
N. Masetti,$^{4,5}$
L. Natalucci,$^{1}$
P. Ubertini.$^{1}$
\\
$^{1}$Istituto di Astrofisica e Planetologia Spaziali (INAF), Via Fosso del Cavaliere 100, Roma, I-00133, Italy\\
$^{2}$Department of Physics \& Astronomy, University of Southampton, Highfield, Southampton, SO17 1BJ, UK\\
$^3$DCPA Georgia College Milledgeville, USA\\
$^{4}$INAF-Osservatorio di Astrofisica Spaziale e Scienza dello Spazio, via Gobetti 93/3, I-40129 Bologna, Italy\\
$^{5}$Departamento de Ciencias F\'isicas, Universidad Andr\'es Bello, Fern\'andez Concha 700, Las Condes, Santiago, Chile
}

\date{Accepted XXX. Received YYY; in original form ZZZ}

\pubyear{2019}

\begin{document}
\label{firstpage}
\pagerange{\pageref{firstpage}--\pageref{lastpage}}
\maketitle

\begin{abstract}
We report on a recent bright outburst from the new X-ray binary transient MAXI J1631$-$479, observed in January 2019. In particular, we present the 30-200 keV analysis of spectral transitions observed with \integral\//IBIS during its Galactic Plane monitoring program. In the MAXI and BAT monitoring period, we observed two different spectral transitions between the high/soft and low/hard states. The \integral\/ spectrum from data taken soon before the second transition, is best described by a Comptonised thermal component with a temperature of $kT_e\sim$ 30 keV  and a high luminosity value of L$_{2-200\,\mathrm{keV}}\sim$3$\times 10^{38}$\,erg\,s$^{-1}$  (assuming a distance of 8~kpc).
During the second transition, the source shows a hard, power-law spectrum.  The lack of high energy cut-off indicates that the hard X-ray spectrum from MAXI~J1631$-$479 is due to a non-thermal emission. Inverse Compton scattering of soft X-ray  photons from a non-thermal  or hybrid thermal/non-thermal electron distribution can explain the observed X-ray spectrum although a contribution to the hard X-ray emission from a jet cannot be determined at this stage.
The outburst evolution in the hardness-intensity  diagram, the spectral characteristics and the rise and decay times of the outburst are suggesting this system is a black hole candidate.

\end{abstract}




\section{Introduction}
In recent years, with the advent of new space and ground-based facilities, numerous efforts have been made to understand the X-ray emission in transient X-ray Binaries, containing either a neutron star (NS) or a black hole (BH). These systems are characterized by transitions between two main spectral states: the high/soft state, with the dominant soft X-ray emission originating from the accretion disc, and the low/hard state, where the dominant hard X-ray emission arises from the Inverse Compton scattering of soft thermal photons by hot electrons in the corona (\cite{done07}). 
The evolution of the spectral and timing properties are crucial to understand the accretion-ejection connection during an outburst for both BH and NS systems.  The evolution of an outburst is well described in the  hardness-intensity diagram, which represents a useful tool to investigate on the phenomenological connections between the spectral-timing states and the outflows modes (\cite{gardenier18,belloni18,belloni16,fender16,fender12,done07,fender04a,fender04}, and references therein).
Although the geometry of these systems is fairly well established, the jet contribution to the high energy emission in the hard state is still unclear.

\begin{figure*}
   \includegraphics[angle=-90,width=8.5 cm]{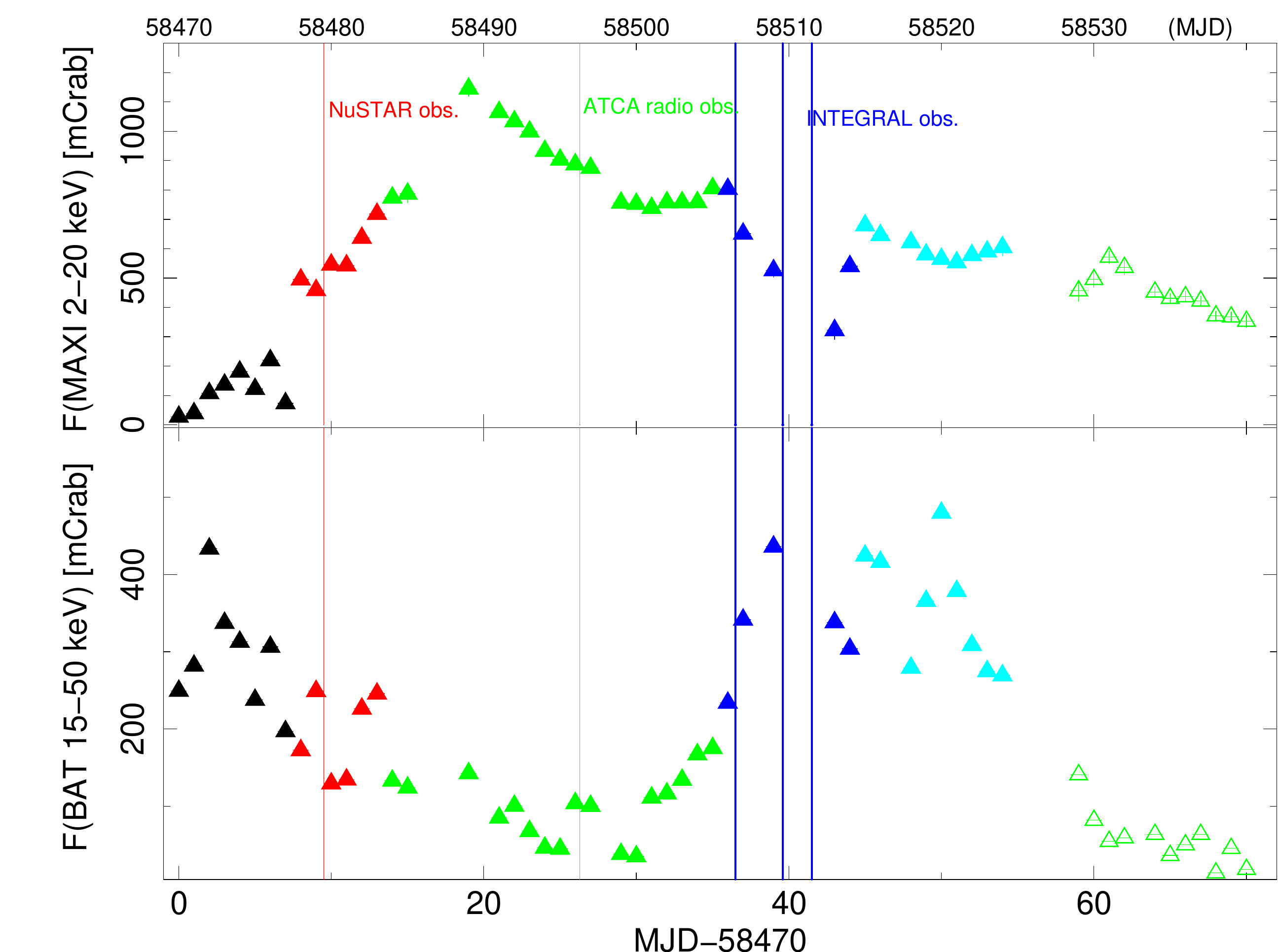}
  \includegraphics[angle=-90,width=8.5cm]{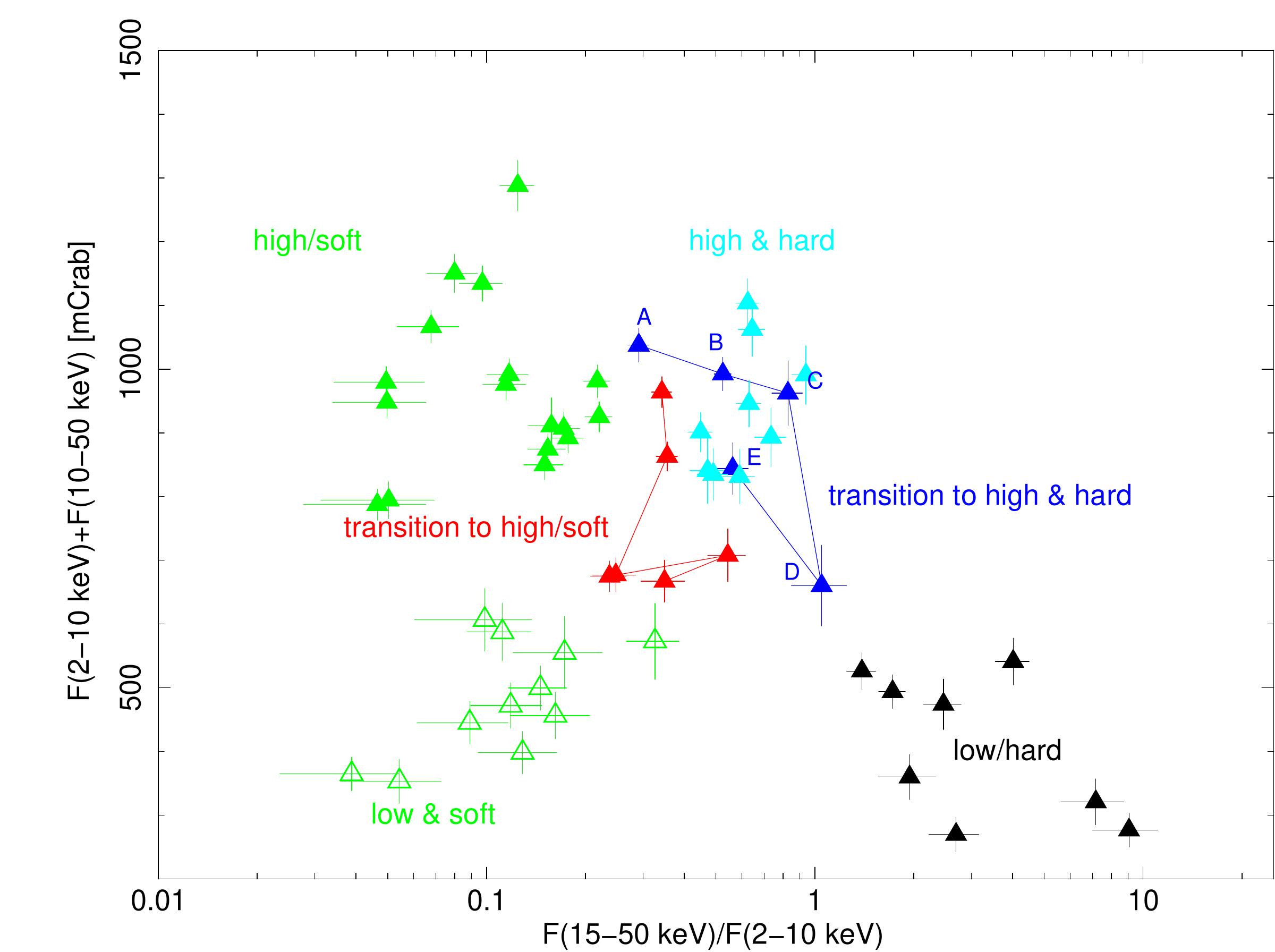}
   \caption{
   {\em Left panel}: the MAXI (upper) and Swift/BAT (lower) light curves of MAXI J1631$-$479, both with 1 day binning.  Vertical lines indicate the time of the \nus\/, ATCA and \integral\/ observations. See text for details.
   {\em Right panel}: the hardness-intensity diagram with each point corresponding to 1 day: on the horizontal axis we show the ratio  $F_{(15-50\,\mathrm{keV})}^{BAT}/F_{(2-10\,\mathrm{keV})}^{MAXI}$ and in the vertical axis the total flux $F_{(15-50\,\mathrm{keV})}^{BAT}+F_{(2-10\,\mathrm{keV})}^{MAXI}$. Colored points indicate the different time intervals. Capitals letters indicate the time evolution of the hardness-intensity data during this transition, with A being the starting of the transition and E the end.}
   \label{fig1}
   \end{figure*}
 
A new outburst from \source\/ was reported on 2018 December 21 by the \maxi\/ nova alert system reporting a bright hard {\it X}-ray outburst in the Norma region \cite[][]{kobayashi18}. The {\it X}-ray flux was F$_{4-10\,\mathrm{keV}}$=209$\pm$27 mCrab at the inferred position of R.A. (J2000)= 247.770 deg., DEC (J2000) = -47.920 deg., with a 90\% confidence error radius of 10.2$'$. 
Confirmation of \source\/ as a new {\it X}-ray transient source was provided by \nus\/ observations performed on 2018 December 28 at the refined position is R.A. (J2000) = 16:31:13.4, DEC (J2000) = -47:48:18 with an uncertainty of 15\arcsec.

On 2019 January 13 a radio counterpart for \source\/ was detected by  the Australia Telescope Compact Array (ATCA) observations at central frequencies of 5.5 and 9 GHz \cite[][]{russel19}. A clear identification of an optical counterpart is missing, consistently with the high column density found in the \nus\/ observation.  Indeed,  the position of \source\/ was observed with KMTNet telescopes \cite[]{kim16},  with the iTelescope.Net T17 and with the 1m telescope of the CHILESCOPE observatory \cite[]{kong19} and no new optical source was clear found at the position of the radio detection \cite[][]{shin19}. 

 \integral\/ started Norma region observations as part of the Galactic Plane Scan in revolution 2048 (2019 January 21- 23) and \source\/ was clearly detected by {\it IBIS}/ISGRI  \cite[][]{onori19}. The position of the source in the 22--60 keV energy band was R.A. (J2000)=247.814 deg, DEC(J2000)=-47.800 deg with an error radius of 0.5$'$ and a flux F$_{22-60\,\mathrm{keV}}$= 265.2$\pm$3.7 mCrab. Here, we report the results from our analysis of the {\it IBIS}/ISGRI data up to 200 keV.


\section{Observations and Data analysis}
 \label{sect_obs}

\source\/ has been monitored at high energies by \integral\/ during three revolutions: 2048, 2049 and 2050, starting from  2019-01-21T13:30 UTC (58504.563 MJD), 2019-01-24T02:47 UTC (58507.117 MJD) and 2019-01-26T19:12 UTC (58509.801 MJD), respectively.\\
The \integral\//IBIS \cite[]{ubertini03} consolidated data for these observations were  processed using the standard Off-line Scientific Analysis (\texttt{osa} v11.0) software, released by the \integral\/ Science Data Centre \cite[]{courvoisier03}. 
This software was used to obtain both spectra and light curves of the source in different common time intervals and 
a systematic error of 2\% was included.
\source\/ was out of the JEM-X field of view and this precludes INTEGRAL detection at
lower energies (<30 keV).

The Neil Gehrels {\it Swift} observatory \cite[]{gehrels04} collected data from this source in January 2019.  We used the results from {\it BAT}\/\cite[]{Krimm13} observations, while the XRT \cite[]{burrows04} data are not simultaneous with the INTEGRAL ones. In view of  the strong variability of the source, a common INTEGRAL-Swift spectral study in a broad energy band is not possible. 

\source\/ was also observed by the Monitor of All-sky X--ray Image \cite[][]{Matsuoka09}, and their data have been used in our analysis. 
 
We use \texttt{xspec} v12.10.1 \cite[]{Arnaud96} in order to fit each spectral state in the 30-200 keV energy range. 

\section{Spectral Analysis Results}
\label{analisis}
In order to study the hard X-ray spectral properties of \source, we analyze the \integral\//IBIS data taken during the three \integral\/ revolutions, over the energy range 30-200 keV.

The time evolution and the respective behavior of the different spectral states are shown in
Figure \ref{fig1}, where the 1-day binned light curves from MAXI\footnote{{\em  http://\maxi\/.riken.jp/top/slist.html}} and {\it Swift}/BAT\footnote{{\em https://swift.gsfc.nasa.gov/results/transients}} are reported. The variation of the source along the outburts is clearly visible from these lightcurves.
The epochs of the \nus\/ (red line), ATCA (green line) and \integral\/ (blue line) observations are shown for comparison.    
The flux conversion into mCrab was obtained with the following factors: 1mCrab$\sim$0.00022 ct\,cm$^{-2}$\,s$^{-1}$ and 1Crab$\sim$3.8 ph\,cm$^{-2}$\,s$^{-1}$ for \textit{Swift}/{\it BAT}~$^{1}$ and \maxi\/~$^{2}$, respectively.

Although the source is within the {\it BAT}\/ confusion radius of another X-ray transient, namely AX J1631.9$-$4752, the \maxi\/ light curve is not contaminated significantly. Indeed, during the monitoring period, while  \source\/ is detected at high flux levels, AX J1631.9$-$4752 is continuously detected at low level by \textit{Swift}/{\it BAT}. The lack of contamination is confirmed by the \integral\//IBIS map, where the emission from AX J1631.9$-$4752 is lower than $\sim$10 mCrab (3$\sigma$ upper limit) in the 30-50 keV energy range.

We performed an analysis for spectral variability by plotting the hardness versus total emission in two energy bands. Figure \ref{fig1} (right panel) shows the hardness-intensity diagram for the \maxi\/ and \textit{Swift}/{\it BAT} observations. The hardness is derived using the ratio between the fluxes in these two X-ray bands F$_{(15-50\,\mathrm{keV})}^{BAT}$/F$_{(2-10\,\mathrm{keV})}^{MAXI}$ and is plotted versus the total flux F$_{(15-50\,\mathrm{keV})}^{BAT}$+F$_{(2-10\,\mathrm{keV})}^{MAXI}$. Each point corresponds to 1 day of time integration and only epochs for which both soft and hard X-ray data are available have been used.   

The different colored points in the right panel of Figure \ref{fig1} represent six regions we have selected from the temporal intervals. These regions correspond to the different spectral states identified in the \maxi\/ and \textit{Swift}/{\it BAT} light curves, as shown in the left panel of Figure \ref{fig1}:
 \begin{itemize}
  \item $1^{st} epoch$: from MJD 58470 to 58478, the source shows a low flux level at soft energy and a high flux level in the hard X-ray band. This is typical behavior for the low/hard state (black points).
  \item $2^{nd} epoch$: from MJD 58479 to 58485, the source shows an increasing soft flux and a decline in hard X-ray, indicating a transition from the low/hard state to a high/soft state (red points). 
  \item $3^{rd} epoch$: from MJD 58486 to 58506, the source shows a high soft flux level and is low in hard X-rays, indicative of a standard high/soft state (green points).
  \item $4^{th} epoch$: from MJD 58507 to 58514, the source shows unusual variability during this period. A decline in the soft X-ray emission is observed for three days followed by a sudden increase lasting for two--three days. Unfortunately, no soft X-ray data are available for the two days in between this change in the emission. In the same period, the source experienced an increase in hard X-rays  corresponding to a soft X-ray decline, followed by 
  a slight drop in hard flux with a corresponding enhancement in soft X-rays. This behaviour suggests that during the first three days the source started a transition from the high/soft state to a low/hard state. Afterwards it suddenly reversed this trend towards a new state (blue points).     
  \item $5^{th} epoch$: from MJD 58515 to 58522, the source shows a high flux level in both soft and hard bands. This suggests that in this epoch \source\/  can be considered being in a intermediate state (light blue points),  where the X-ray emission can be explained with hybrid (thermal and non-thermal) Comptonization models.
   \item $6^{th} epoch$: from MJD 58529 to 58540, the source transits towards a soft state at low luminosity level (green empty points).
\end{itemize}

 \begin{figure*}
   \includegraphics[angle=-90,width=7.7cm]{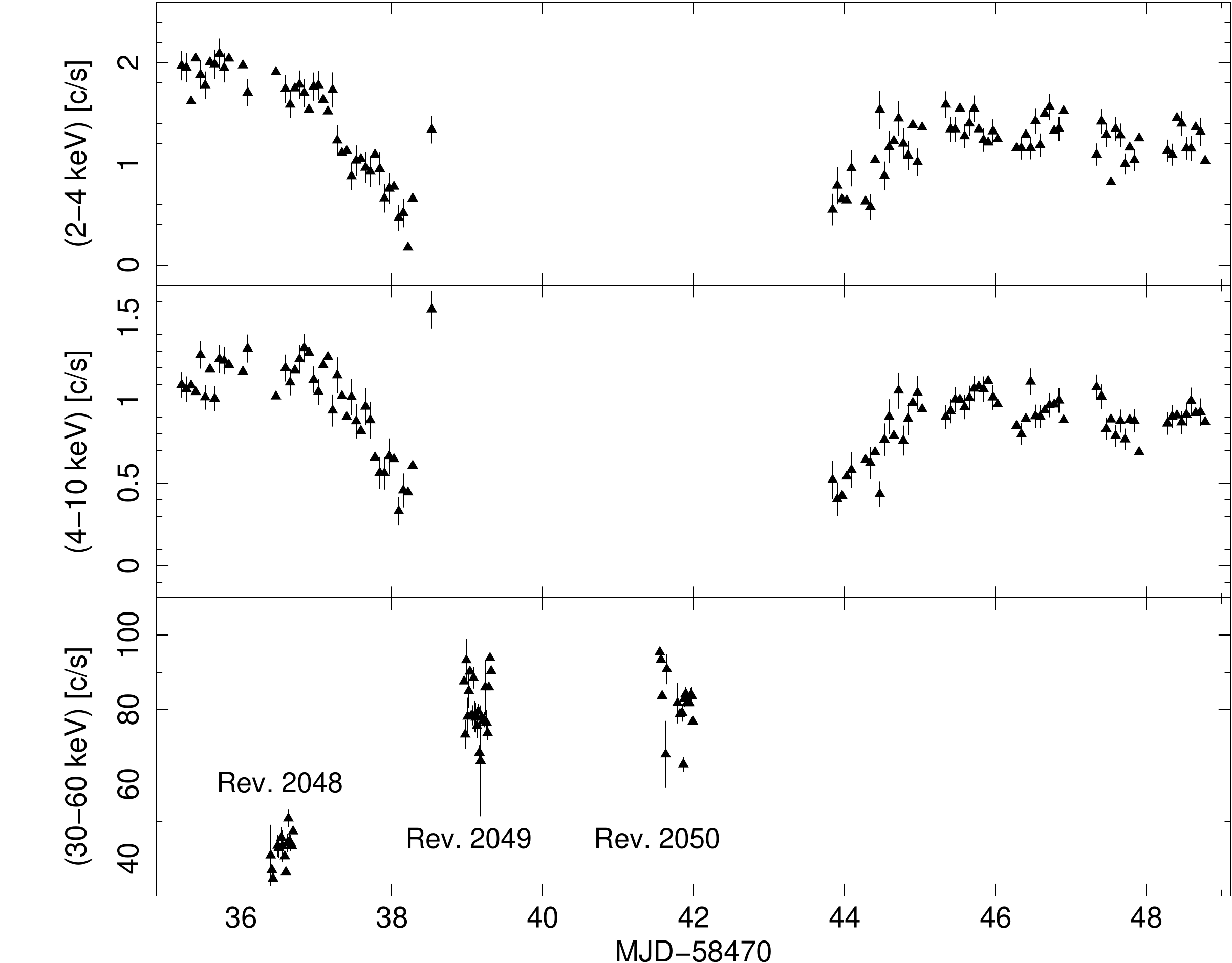}
   \includegraphics[angle=-90,width=9.9cm]{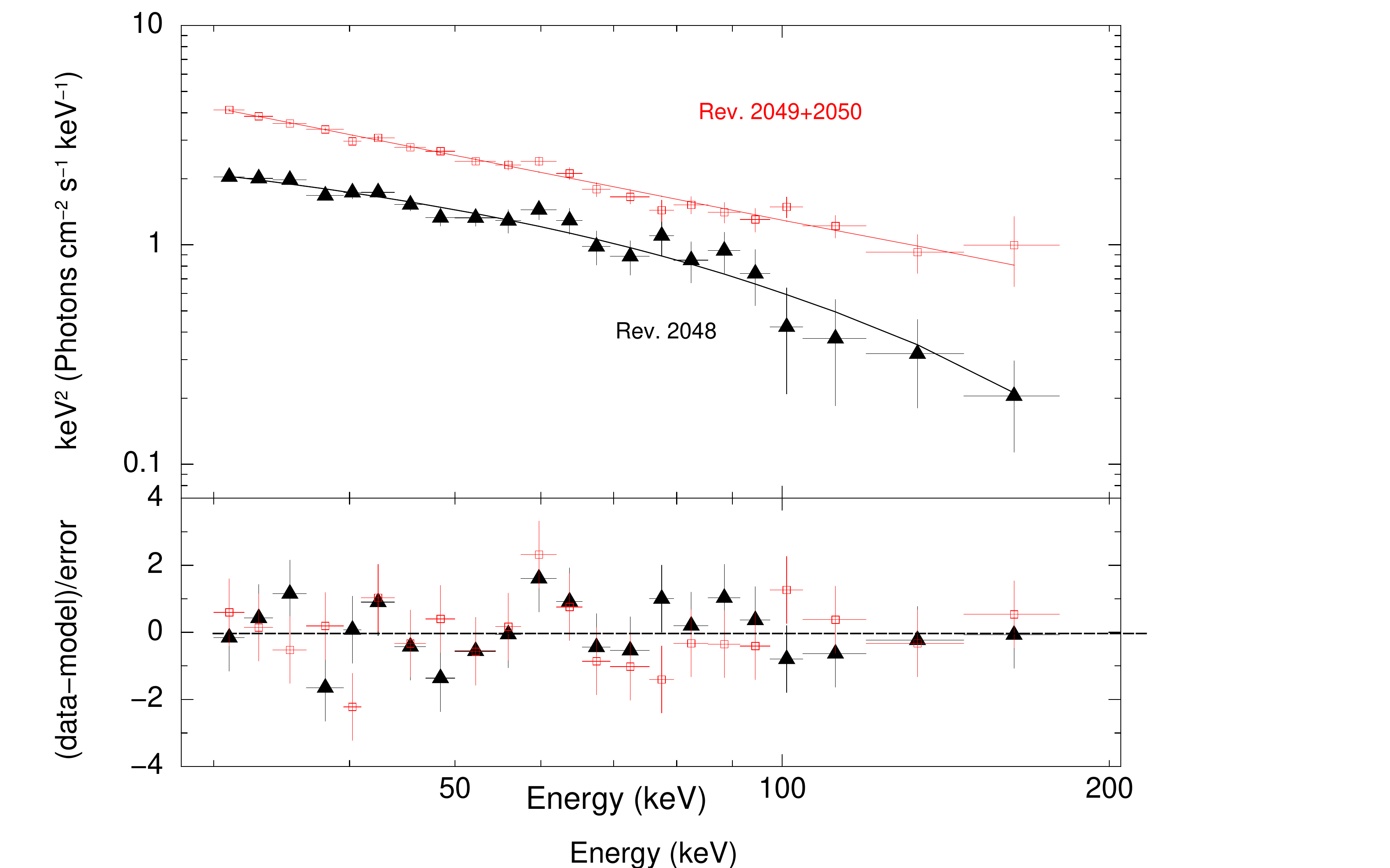}
   \caption{
   {\em Left panel}: Zoomed-in light curves during the \integral\/ observation period, with the top (2-4 keV) and middle (4-10 keV) \maxi\/ light curves (in 6h bins, data from {\em http://maxi.riken.jp/top/slist.html}) compared with the bottom (30-60 keV) IBIS light curve (in 2000 s bins) over the three \integral\/ revolutions. 
   {\em Right panel}: \integral\//IBIS unfolded spectra and residuals in sigma during revolution 2048 (in black) and during revolutions 2049 and 2050 (in red). Model is a cut-off power-law for the first data set and a simple power-law for the second one.
   }
   \label{fig2}
   \end{figure*}
   
The IBIS/\integral\/ observations occurred during the $4^{th}$ epoch and fall in the region of the blue points in the right panel of Figure \ref{fig1} and marked as blue vertical lines in the left panel, for comparison.
The IBIS observations carried out during revolution 2048 fall close in time to the first \maxi\/ blue point (point A in figure \ref{fig1}, right panel). No \maxi\/ and {\it BAT}\/ simultaneous data are available for the second and third IBIS/\textit{INTEGRAL} observations.

The left panel of Figure \ref{fig2} shows 
the 2--4 keV and 4--10 keV MAXI light curves (bin time 6h), together with the 30--60 keV IBIS light curve (bin time 2000 s).
The variation in the source spectral state during the \integral\/ observation is clearly visible.  In particular, during the first \integral\//IBIS observation (revolution 2048) the source is in a high/soft state, however during the subsequent observations (revolutions 2049 and 2050) the source flux  increased by a factor $\sim$2 in the 30--60 keV energy range, indicating the occurrence of the transition to a low/hard state.

We performed spectral analysis for the three \integral\/ revolutions separately.

First, we attempted to fit the spectra using a simple power law model. For the spectrum of revolution 2048 we obtain a spectral index of $3.1\pm0.1$ and a 
$\chi^{2}/$d.o.f. of 34/20,
with residuals at high energies. Then we modelled these data with a cut-off power-law, which resulted in a 
best-fit with $\chi^{2}/$d.o.f.=25/19. In this case, we obtain a spectral index $\Gamma=2.1_{-0.5}^{+0.6}$, E$_{\rm cutoff}=62_{-22}^{+20}$ keV and a high energy flux of F$_{30-200\,\mathrm{keV}}$=(2.8$\pm$0.1)$\times$10$^{-9}$ \cgs.  However we remark that by using the latter model the fit improvement is only marginal. Indeed, applying the appropriate F-test statistics for model comparison (Press et al. 2007; see also Orlandini et al. 2012), we obtain an improvement probability of only about 28\%.

In order to obtain the physical parameters of the source, we also fit this dataset using a model describing the Comptonization of soft photons in a hot plasma \cite[\texttt{comptt} in \texttt{xspec},][]{titarchuk94}. The resulting Comptonized plasma has a temperature  of $\sim$29 keV and optical depth $\tau\sim$ 0.7, with the input soft photon Wien temperature fixed at the default value (0.1 keV), not affecting the high energy spectrum. The best fit parameters are listed in Table \ref{fit}. In the right panel of  Figure \ref{fig2} the spectrum of revolution 2048, together with residuals with respect to the cut-off power-law model are shown (black filled triangles). 

When fitting the IBIS data of revolutions 2049 and 2050 separately, we obtain similar results, both in flux and in spectral shape. This is also evident from Figure \ref{fig1} (right panel), where IBIS observations fall in the same hardness region (between points C and D). 
Using a simple power law model, 
we obtain a spectral index of $3.0\pm0.2$, a flux of F$_{30-200\,\mathrm{keV}}$=(5.9$\pm$0.2)$\times$10$^{-9}$ \cgs and a $\chi^{2}/$d.o.f.	of 20/20 for revolution 2049 and for revolution 2050  ($\Gamma=3.0\pm0.2$, a flux of F$_{ 30-200\,\mathrm{keV}}$=(5.9$\pm$0.2)$\times$10$^{-9}$ \cgs and a $\chi^{2}/$d.o.f.	of 20/20). No high energy cutoff is required for these data.
Taking into account the spectral similarity, similar flux levels and hardness, data from revolutions 2049 and 2050 have been combined to improve the statistical  quality  of  the  spectra  and  to better  constrain the  physical  parameters, as reported in Table \ref{fit}. The resulting spectrum together with the residuals with respect to the power law model are shown in Figure \ref{fig2} with red squares (right panel).

\begin{table}
\caption{Spectral results of the IBIS/INTEGRAL observations.}
\label{fit}
\centering
\begin{tabular}{llll}
\hline\hline
Model				        &Parameter			     	&Rev. 2048				&Rev. 2049+2050			\\
\hline			
		    	                &		                        &\multicolumn{2}{c}{IBIS Exposure} \\
		    			    	&		                        &	13 ks           &36 ks              \\
\hline
&&&\\
Power Law                   &$\Gamma$ 				    &3.1$\pm$0.1			&2.98$\pm$0.04\\
                            &F$^a_{30-200\,\mathrm{keV}}$	    &3.0$\pm$0.1	        &5.9$\pm$0.1	          \\
                            &$\chi^{2}/$d.o.f.	&34/20				    &20/20\\
\hline
&&&\\
Cutoff                      &  $\Gamma$ 				&2.1$_{-0.5}^{+0.6}$    &...\\
Power law                   &E$_{\rm cutoff}$ (keV)		& 62$_{-22}^{+20}$	    &...\\     
                            &F$^a_{30-200\,\mathrm{keV}}$	    &2.8$\pm$0.1	        &... \\
                            &$\chi^{2}/$d.o.f.	&25/19      			&...\\
\hline
&&&\\
Comptt                      & $kT_e$ (keV)              &29$_{-10}^{+43}$         & ...\\
                            &$\tau$	                    &0.7$\pm0.4$	          &...	\\
                            &F$^a_{30-200\,\mathrm{keV}}$	    &2.9$\pm$0.1	          &...	\\
                            &$\chi^{2}/$d.o.f.	&25/19		  	  &...\\
\hline
&&&\\
\end{tabular}
$^a$ Flux in units of 10$^{-9}$ \cgs\\
\end{table}

\section{Discussion} 
\label{sect_disc}
\source\/ belongs to a class of transient systems showing spectral state transitions. 

When it starts to transit towards a hard state \cite[][]{negoro19} 
 the \integral\//IBIS observation shows that the source had a high X-ray flux F$_{30-200\,\mathrm{keV}}\sim$3$\times 10^{-9}$ erg\,cm$^{-2}$\,s$^{-1}$and a hard spectrum (blue lines of Figure \ref{fig1}, at $\sim$ 58506 MJD)
and it is dominated by 
a hard X-ray Comptonized component, arising from inverse Compton scattering of soft thermal photons in a hot corona with $kT_e\sim$ 29 keV and $\tau\sim$ 0.7. 
 
This intermediate value of the electron temperature 
confirms that the source is not in a standard  high/soft or low/hard state. 
\citeauthor[]{joinet08} (2008) found a similar corona temperature ($\sim$ 30 keV) when the microquasar GRO J1655--40 transits from low/hard to hard/intermediate state. 
This temperature of the thermal population was also previously observed in GX 339-4 (\cite{motta09}) before the source transits from low/hard to hard/intermediate state. In this case the high energy cutoff disappears near the hard/intermediate-soft/intermediate transition. This behavior is similar to the \source\/ evolution: the high energy cutoff was not detected by \emph{INTEGRAL}/IBIS when it moved towards a soft state  at low luminosity. 
Indeed it is detected up to about 200 keV though the best fit is a simple power law ($\Gamma\sim 3$) suggesting a non-thermal origin for this X-ray emission.
 This behavior continued in later observations, as noted in  Figure \ref{fig1}, where the light blue points show that the source moved towards an intermediate state at a high flux in both soft and hard bands. 

Transitions from low/hard state to high/soft state and then towards an intermediate state with high flux at these energies has previously been observed in 
black hole systems \cite[][]{done07,belloni18}.  The no-detection of a high energy cutoff using X-ray data up to 200 keV suggests that the hard X-ray emission may not be produced by thermal Comptonization. 
The observed spectrum in the intermediate state can be produced by Compton scattering on a non-thermal electron population \cite[][]{gilfanov10}
or by a mixture of thermal and non-thermal features \cite[][]{coppi99,delsanto08}.
The shape of the hysteresis diagram of \source\/  appears similar to the one of the black hole X-ray binary  XTE J1550$-$564 \cite[]{russel10}.
 Moreover, \source\/ showed radio emission at 5.5 and 9 GHz when it was in the high/soft state \cite[][]{russel19} and a similar behaviour was reported for XTE J1550$-$564.
For that system, the authors were able to isolate the non-thermal emission from the jet and demonstrated that 
the synchrotron jet may dominate in hard X-ray when the source fades in the low/hard state, at very low luminosity level. In this case, the jet produces $\sim$ 100\% of the emission  when the soft (3-10 keV) and hard (100-250 keV) fluxes are comparable. 

In the case of \source\/, Figure \ref{fig2} shows that X-ray power-law component is observed when the soft and hard fluxes are comparable, although the energy ranges are slightly different from those used for XTE J1550$-$564.

Unfortunately, for this source neither the distance nor the mass of the central object are known. In order to estimate the possible luminosity range of \source\ , we used the extreme values of distances and masses known so far for X-ray binaries. In particular, the distances span from 1 kpc \cite[][]{McC86} to 27 kpc \cite[][]{casares04}, while the black hole masses can assume values from 4 M$_{\sun\/}$ \cite[][]{ozel1O} to 15.65 M$_{\sun\/}$ \cite[][]{McC86,orosz07}.  Using realistic values for a LMXB, the luminosity of \source\/ during the intermediate state can vary from 5$\times10^{-1}$ L$_{Edd}$ (distance=20 kpc and mass=4 M$_{\sun\/}$) to 3.5$\times10^{-4}$ L$_{Edd}$ (distance=2 kpc and mass=15.65 M$_{\sun\/}$). Thus, as the emission of the jet in X-ray band can be dominant when the luminosity is between L$\sim$2$\times 10^{-3}$ L$_{Edd}$ and L$\sim$2$\times 10^{-4}$ L$_{Edd}$ \cite[]{russel10}, the wide range of possible values derived for the luminosity of \source\/ do not allow us to exclude a contribute of the jet to the X-ray emission in \source\/. 

The lack of a high energy cut-off indicates that the X-ray emission from \source\/ has a non-thermal  origin that can be produced either by  Comptonization of non-thermal electron population or by a hybrid thermal/non-thermal electron distribution. 
 
The source outburst evolution  and the typical times of the outburst suggest the BH nature of this source. Indeed,
  assuming  the time spent to reach from 10\% to 90\% of the flux peak as the rise time and the reverse for the decay time,  we obtained $\tau_{rise}\sim$15 days and $\tau_{decay}\geq$50 days though it seems the outburst has not yet ended  (see Fig 1, left panel). These values are compatible with decay and rise times for BH and incompatible with much faster evolution times for NS, reported in the statistical study of Yan and Yu (2015).
 This result is in agreement with previous studies performed at different energy bands during the same outburst.  
 When \source\/ transits from a low/hard to high/soft state  (red line of Figure \ref{fig1}, $\sim$58480 MJD) the \nus\/ 3--79 keV spectrum  is well modelled by a disk-blackbody with a temperature of $\sim$1.12 keV, a power-law with a photon index of $\sim$2.39 
and iron K$\upalpha$ emission line with an equivalent width of $\sim$ 90 eV
\cite[][]{miyasaka18}.
 Based on these spectral characteristics the authors  suggested the system as a black hole binary in the high/soft state with most of the flux in the soft X-ray band (F$_{2-10\,\mathrm{keV}}$=1.7$\times$10$^{-8}$ \cgs\/and F$_{3-79\,\mathrm{keV}}$=1.8$\times$10$^{-8}$ \cgs). 
 \citeauthor{russel19} (2019) also report a similar conclusion on the source nature by using the radio observation performed on 2019 January 13 (green line of Figure \ref{fig1}, $\sim$58496 MJD). A radio counterpart of \source\/ was detected by Australia Telescope Compact Array (ATCA) with a  flux density of  F$_{5.5\,\mathrm{GHz}}$=(630 $\pm$ 50) $\upmu$Jy 
 and a radio luminosity L$_{5\,\mathrm{GHz}}$=(9.5 $\pm$ 0.8)$\times$10$^{28}$~(d/5 kpc)$^{2}$ erg\,s$^{-1}$. 

This radio emission could indicates  the presence of
a compact jet, as it was detected during the transition from high/soft to low/hard state, i.e. when synchroton jet emission could became dominant at X-ray wavelengths and at low luminosities.

\section*{Acknowledgements}
We acknowledge  the ASI financial/programmatic support via  ASI-INAF agreement number 2013-025.R1 and  ASI/INAF n.2017-14-H.0.\\
F.O. acknowledge the support of the H2020 European Hemera program, grant agreement No 730970.\\ 
This research has made use of the \maxi\/ data provided by RIKEN, JAXA and the \maxi\/ team.\\ 
{\it Swift}/{\it BAT}\/transient monitor results provided by the {\it Swift}/{\it BAT}\/team.\\
P.A.C. is grateful to the Leverhulme Trust for the award of an Emeritus Fellowship.\\
Finally, the authors thank the anonymous referee for the valuable comments which helped to improve the manuscript.





\bibliographystyle{mnras}
\bibliography{MAXI} 







\label{lastpage}
\end{document}